\documentclass[traditabstract]{aa} % for the abstract without structuration % (traditional abstract) 
\usepackage{graphicx}
\usepackage{txfonts}
\usepackage{natbib}%[comma]
\usepackage{verbatim}
\usepackage{comment}
\usepackage{color}
\usepackage{bm}
\usepackage{txfonts}
\usepackage{float}
\def\nodata{\multicolumn{1}{c}{$\cdots$}}
\newcommand{\so}{$\sigma$~Orionis~}

\newcommand{\mj}{\,M$_{\rm Jup}$~}
\newcommand{\ms}{\,M$_{\odot}$}
\bibpunct{(}{)}{;}{a}{}{,} % to follow the A&A style

\begin{document}
\title{A new free-floating planet in the Upper Scorpius association}
\subtitle{}
\author{K. Pe\~na  
Ram\'irez\inst{1,2} \and V. J. S. B\'ejar\inst{3,4} \and M. R. Zapatero Osorio\inst{5} }
\institute{Instituto de Astrof\'isica. Pontificia Universidad Cat\'olica de Chile (IA-PUC), E-7820436 Santiago, Chile. \email{kpena@astro.puc.cl} \and Millennium Institute of Astrophysics, Santiago, Chile \and Instituto de Astrof\'isica de Canarias (IAC), E-38205 La Laguna, Tenerife, Spain. \and Universidad de La Laguna, Tenerife, Spain.\and Centro de Astrobiolog\'ia (CSIC-INTA), E-28850 Torrej\'on de Ardoz, Madrid, Spain. \\}
\date{Received 23 September 2015 / Accepted 17 November 2015}

% \abstract{}{}{}{}{} 
\abstract
% context heading (optional)
%{.}
% aims heading (mandatory)
%{.}
% methods heading (mandatory)
%{.}
% results heading (mandatory)
%{.}
% conclusions heading (optional)
%{}
{We report on a deep photometric survey covering an area of 1.17 deg$^2$ in the young Upper Scorpius stellar association using VIMOS $Iz$ and UKIDSS $ZJHK$ data taking several years apart. The search for the least massive population of Upper Scorpius ($\sim$5--10\,Myr, 145\,pc) is performed on the basis of various optical and infrared color-color and color-magnitude diagrams, including {\sl WISE} photometry, in the magnitude interval $J$=14.5-19\,mag (completeness), which corresponds to substellar masses from 0.028 through 0.004~\ms~at the age and distance of Upper Scorpius. We also present the proper motion analysis of the photometric candidates, finding that two objects successfully pass all photometric and astrometric criteria for membership in the young stellar association. One of them, USco\,J155150.2$-$213457, is a new discovery. We obtained low resolution, near-infrared spectroscopy ($R\sim$450, 0.85--2.35\,$\mu$m) of this new finding using the FIRE instrument. We confirmed its low-gravity atmosphere expected for an Upper Scorpius member (weak alkaline lines, strong VO absorption, peaked $H$-band pseudocontinuum). By comparison with spectroscopic standards, we derive a spectral  
type of L6\,$\pm$\,1, and estimate a mass of $\approx$0.008--0.010~\ms~for USco\,J155150.2$-$213457. The colors and spectral slope of this object resemble those of other young, cool members of Upper Scorpius and $\sigma$ Orionis ($\sim$3\,Myr) and field, high gravity dwarfs of related classification in contrast with the very red indices of field, low gravity, L-type dwarfs of intermediate age. USco\,J155150.2$-$213457, which does not show infrared flux excesses up to 4.5\,$\mu$m, becomes one of the least massive and latest type objects known in the entire Upper Scorpius stellar association.}

\maketitle

%%%%%%%%%%%%%%%%%%%%%%%%%%%
\section{Introduction \label{intro}}
The shape of the initial brown dwarf and planetary mass function and the minimum mass for the collapse and fragmentation of clouds are crucial topics to understand the dominant substellar formation process. Since substellar objects are significantly brighter and warmer at very young ages, e.g., less than 10 Myr  \citep{chabrier00model}, the detection of sources with a few Jupiter masses is possible by exploring nearby star-forming regions. Deep searches for the least massive population of these regions may shed light on the aforementioned topics.

The proximity and youth of the Upper Scorpius association (USco from now on) make this region more than suitable to performing searches for members in the substellar regime. The entire USco covers a vast area of more than 200\,deg$^2$ on the sky \citep{slesnick08} and is located at a mean distance of 145$\pm$2\,pc \citep{dezeeuw99}. We carried out a deep photometric survey in a zone of low foreground extinction (A$_{\rm v}<$1\,mag; \citealt[and references therein]{preibisch98}). Regarding the USco age, previous studies suggested that the association is $\sim$5\,Myr old on the basis of the location of B-type stars in the Hertzsprung-Russell diagram and and their comparison with evolutionary tracks \citep{degeus89,preibisch99,preibisch02}. Recently, \citet{herczeg15} estimated an age of $\sim$4\,Myr for the low mass members of the USco association calculated from models that reproduce the lithium depletion boundary of various young star clusters and stellar moving groups. This is somehow younger than the USco age derived by \citet{pecaut12}. These authors estimated an age of 11\,$\pm$\,2\,Myr for intermediate and high mass USco members, including the M-type supergiant star Antares. The newly discovered eclipsing binaries in USco (UScoCTIO\,5 and EPIC\,203868608,   \citealt{kraus15,david15}) also support ages close to the 10\,Myr for this stellar association. In addition, the high proper motion of the USco association ($\mu$\,=\,26.7\,$\pm$\,2.5\,mas\,yr$^{-1}$, \citealt{zacharias04}) is beneficial for the unambiguous identification of its true members once the astrometric and photometric studies are conveniently combined.

Here, we present an exploration extending over 1.17\,deg$^2$ in the USco association. We used deep photometric data covering the wavelength range 0.8--3.4\,$\mu$m. Our goal was to define the USco sequence of members with masses ranging from $\sim$0.025 through 0.004\ms\,within completeness. We also performed the spectroscopic follow--up of the faintest and least massive candidate found in our survey. The observational dataset is described in Section\,\ref{data}. The photometric and astrometric selection of USco member candidates and a discussion on possible field contaminants appear in Section\,\ref{selec}. Section\,\ref{anal_spectrum} introduces the spectroscopic data analysis of our faintest candidate. Mass values estimated from theoretical isochrones are presented in Section\,\ref{umodels}. In Section\,\ref{final} we put our search in the context of other explorations carried out in the USco region, and we discuss the implications of our findings for understanding the USco mass function. Our conclusions are given in Section\,\ref{conc}.

%%%%%%%%%%%%%%%%%%%%%%%%%%%%%%%%%%

\section{Observational data}\label{data}
\subsection{Optical and near--infrared imaging}
\begin{figure}
\centering
\includegraphics[width=0.5\textwidth]{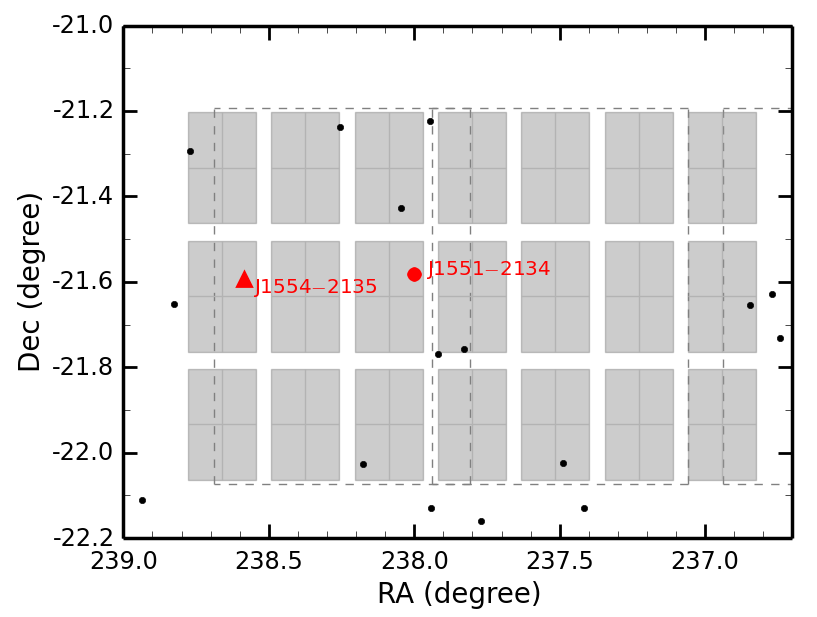}\\
\caption{Distribution of the 21 VIMOS pointings (gray squares) in the right ascension---declination diagram. The UKIDSS coverage overlapping with the VIMOS survey is indicated by the dashed gray lines \citep{lodieu06, lodieu11}. The red symbols stand for the two final USco member candidates -- labelled with their abridged names -- selected from our combined photometric and astrometric study. Black dots represent the known USco member candidates and confirmed members \citep{lodieu06,lodieu11,bejar09,dawson14} that fall within the VIMOS region and its surroundings. 
\label{upoint}}
\end{figure}

Our survey, which covered an area of 1.17 deg$^2$ in USco, was carried out with the VIsible MultiObject Spectrograph (VIMOS; \citealt{vimos}) installed on the Nasmyth A focus of 8-m Unit~3 (Melipal) of the Very Large Telescope (VLT) array sited on Cerro Paranal, Chile. In imaging mode, VIMOS field of view is 7\,$\times$\,8\,arcmin$^2$ per pointing. It uses a squared mosaic of four detectors (2048\,$\times$\,2440\,pix$^2$) separated by a gap of 2\arcmin. The instrumental pixel scale is 0\farcs205. A total of 21 VIMOS pointings were acquired using the $I$ and $z$ filters\footnote{Central wavelengths and bandwidths ($\mu$m) as follows: $I_c\,=\,0.823$, $z_c\,=\,0.914$;  $\Delta I\,=\,0.211$, $\Delta z\,=\,0.185$. The transmission of the VIMOS filters slightly changes from detector to detector. The values above correspond to the average of the four detectors. In this paper, all quoted magnitudes are given in the Vega system.} and were designed to overlap with the area explored by the UKIRT Infrared Deep Sky Survey\footnote{UKIDSS uses the UKIRT Wide Field Camera (WFCAM; \cite{casali07}) and a photometric system described in \cite{hewett06}. The pipeline processing and science archive are described in \cite{hambly08}. The UKIDSS broadband $ZJHK$ data are directly calibrated from 2MASS point sources \citep{hodgkin09}.} (UKIDSS, \citealt{lawrence07}). This gave us access to complementary photometric data in the $ZYJHK$ bands, i.e., from 0.823 through 2.2 $\mu$m. Figure~\ref{upoint} shows the distribution of our VIMOS pointings in the north-east part of the USco region. Each pointing consisted of three exposures per filter with individual integration times of 160\,s ($I$) and 105\,s ($z$). The total exposure times were 8 min ($I$) and 5.2\,min ($z$) per pointing. In Table~\ref{u:log}, we provide the central coordinates of all VIMOS pointings, the observing filters and dates, the exposure times, and the average seeing conditions as measured from the final reduced images.

The four detectors of VIMOS were treated separately for a proper processing of the raw data. Raw images were reduced using the CCDRED routine within the IRAF\footnote{The Image Reduction and Analysis Facility (IRAF) is distributed by National Optical Astronomy Observatories, which is operated by the Association of Universities for research in Astronomy, Inc., under contract to the National Science Foundation.} environment. We applied a bias correction using the overscan regions. The flat-field correction was done by generating a super-flat image (one per filter) from the median combination of all VIMOS science images. With these super-flat frames we managed to reduce the fringing pattern, which is particularly strong for the $z$ band \citep{lagerholm12}. All science images were aligned and combined to produce deep frames per pointing and filter. Finally, we cut the vignetted areas from the processed images\footnote{In imaging mode, the VIMOS illuminated area is $\sim$1970\,$\times$\,2300\,pix$^2$. Exact values per detector were taken from VLT$-$MAN$-$ESO$-$14610$-$3509v86$.$1$.$pdf as of May 2015.}. The area of 1.17 deg$^2$ was determined from the cut images.

The photometric analysis was made using DAOPHOT routines within IRAF. First, we automatically identified unresolved sources in the reduced images using the DAOFIND routine. Aperture and point-spread-function (PSF) photometry was performed on the identified sources with an aperture radius of 4\,$\times$\,FWHM. The stellar PSF was defined from the Gaussian fit to about 10--15 isolated stars homogeneously distributed across the frames. Instrumental magnitudes in the $I$-band were converted into observed magnitudes using data from the DEep Near Infrared Survey of the Southern Sky (DENIS; \citealt{epchtein94}). We searched for common sources between DENIS and our data in the DENIS magnitude range $I$\,=\,17--18 mag. \citet{costado05} found that the DENIS $I$ and VIMOS $I_{\rm Cousin}$ magnitudes are quite alike with a negligible difference of $I_{\rm Cousin}-I_{\rm DENIS}$\,=\,0.03\,$\pm$\,0.04\,mag for M and L type sources, which is the spectral type domain of our interest. Therefore, we did not correct the VIMOS $I$-band magnitudes for any color term. Regarding the $z$-band photometry, the VIMOS instrumental magnitudes were calibrated using the $Z$-band data from the UKIDSS Eight Data Release (DR8) using sources in common that have UKIDSS magnitudes in the  interval  $Z$\,=\,16--17.5 mag. The VIMOS $z$ filter is centered at a redder wavelength than the UKIDSS $Z$-band; however, after multiplying the VIMOS $z$ filter by the response function of the detectors, the bandpasses of the two filters become very similar. To further test whether there is a color term between VIMOS $z$ and UKIDSS $Z$ magnitudes, we retrieved the near-infrared spectra of L-type field dwarfs from the catalog by \citet{rayner09}. All spectra were conveniently convolved with the filter transmission profiles and corresponding detectors response functions and integrated to derived the $z$ and $Z$ magnitudes. We found that the resulting magnitudes are alike with negligible differences confirming that no color term is required for calibrating the VIMOS $z$-band. The dispersion of the photometric calibration was determined at $\pm$0.09\,mag for the $I$-band and $\pm$0.04\,mag for the $z$-band. These quantities were quadratically added to the instrumental PSF magnitude errors provided by IRAF. 

\begin{figure}
\centering
\includegraphics[width=0.5\textwidth]{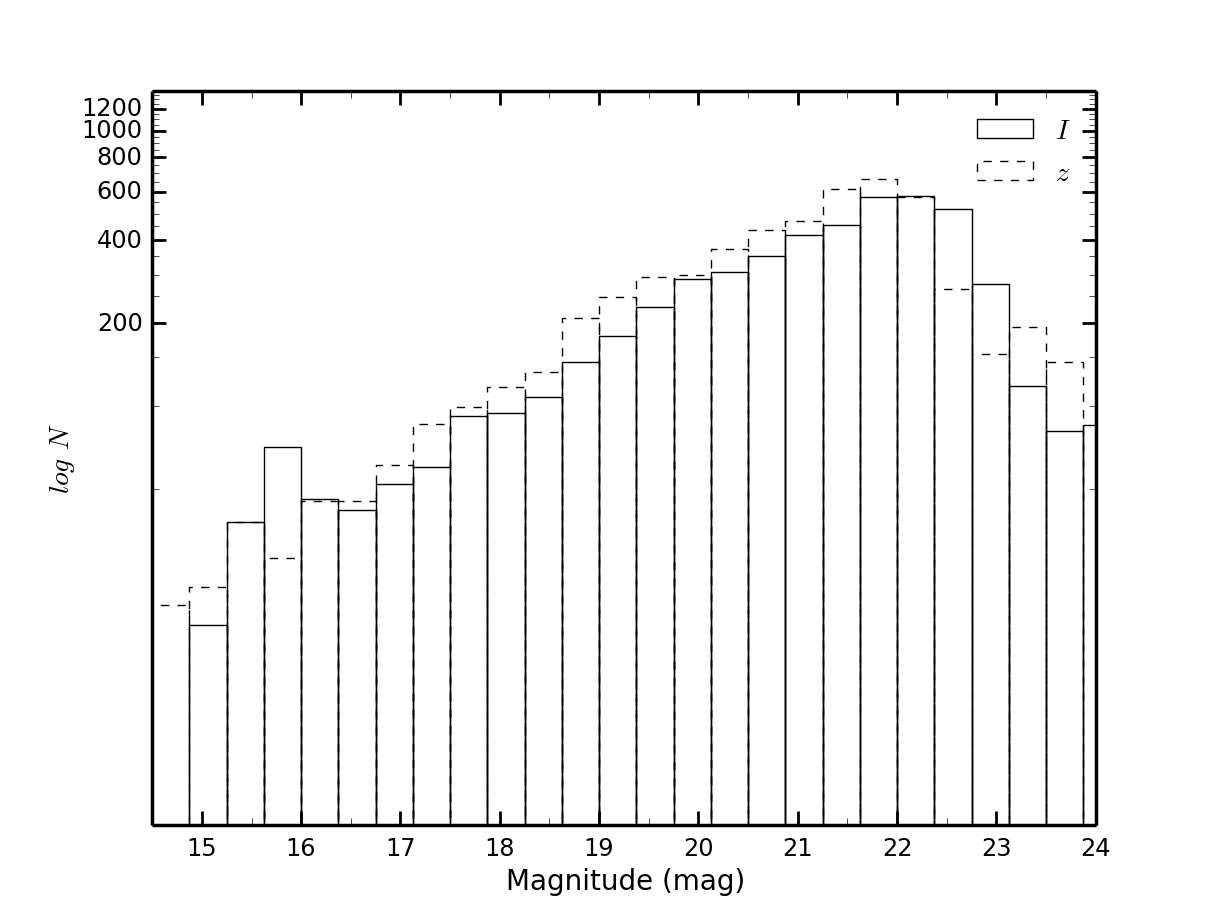}\\
\caption{Source counts as a function of observed VIMOS $I$ and $z$ magnitudes for the fifth pointing, which has the shallowest images and defines the completeness magnitudes of the $Iz$ survey. The bin size is 0.5 mag. 
\label{histogram}}
\end{figure}

We estimated the completeness magnitudes of the VIMOS survey ($Iz$-bands) and the combined VIMOS--UKIDSS exploration ($Iz+ZJHK$ bands, see Section~\ref{cri1}) from the comparison of the total number of identified sources per magnitude interval with an exponential distribution of stars. The exponential fit was obtained for each filter independently using bright to intermediate magnitude sources. The completeness magnitude is defined as the magnitude interval that immediately precedes the magnitude bins displaying a continuous deficit of stars with respect to the exponential prediction. The limiting magnitude is calculated as the magnitude bin at which the total number of sources deviates by $\geq$50\% from the counts of the completeness magnitude bin. Our determined completeness and limiting magnitudes are given in Table~\ref{u:log} and roughly correspond to source detections around the 10-$\sigma$ and 4-$\sigma$ levels, respectively ($\sigma$ is the sky background noise). In short, the USco VIMOS survey is complete down to $z$\,=\,21.7 and $I$\,=\,22.0\,mag. These values correspond to the shallowest images of VIMOS pointing number 5, whose source counts versus observed magnitudes are illustrated in Figure\,\ref{histogram}. The deepest VIMOS pointing (number 11) has images that are $\sim$1.1\,mag fainter in both bands. The limiting magnitudes of the VIMOS survey are $z$\,=\,23.5 and $I$\,=\,23.8\,mag (computed as the mean values of all 21 pointings). Regarding the UKIDSS data and following the same approach, we determined the following completeness and limiting magnitudes: $Z$\,=\,20.8, $Y$\,=\,20.5, $J$\,=\,19.0, $H$\,=\,18.7, and $K$\,=\,18.0\,mag (completeness), and $Z$\,=\,21.5, $Y$\,=\,21.2, $J$\,=\,20.0, $H$\,=\,19.4, and $K$\,=\,18.7\,mag (limiting magnitudes). These quantities agree with the values given by other groups \citep{lodieu06} and with the information provided on the UKIDSS webpage.\footnote{http://surveys.roe.ac.uk/wsa/dr8plus\_release.html as of July 2015.} In view of these numbers, our VIMOS survey is 2\,mag deeper than the UKIDSS data. 

The VIMOS data were astrometrically calibrated using the right ascension and declination coordinates of the 2MASS catalog \citep{skrutskie06}. The internal precision of the astrometric calibration is $\pm$0\farcs2.

\subsection{Near--infrared spectroscopy}
Near-infrared $JHK$ spectroscopy of USco\,J155150.2$-$213457 (hereafter J1551$-$2134, see Section \ref{selec}) was obtained using the Folded-port InfraRed Echellette (FIRE; \citet{simcoe08,simcoe13}) instrument installed at the 6.5-m Baade Telescope, one of the Magellan telescopes sited on Las Campanas Observatory (Chile). FIRE is a near-infrared dual-mode spectrometer (detector pixel size of 0\farcs15) that operates over the 0.82--2.51\,$\mu$m wavelength range. We used the FIRE high-throughput prism mode with a slit width of 0\farcs6. This instrumental configuration yielded a spectral nominal dispersion of 9.25\,\AA\,pix$^{-1}$ and a resolving power of about 450 at the central frequency ($\lambda_c\sim$ 1.66\,$\mu$m). Because of the faint nature of our target, the useful wavelength spectrum coverage is 0.85--2.35\,$\mu$m. Observations of J1551$-$2134 were collected with a seeing of 0\farcs5--0\farcs6 on 2015 May 13.

J1551$-$2134 and a bright reference star ($J$=14.6\,mag) at a separation of 29\farcs6 from the science target were acquired with the $J$-band filter and simultaneously aligned on the nominal 1\arcmin-length slit. Individual 300\,s exposures were obtained with the source at two nod positions separated by 5\farcs0. Both the science target and the reference star were observed in an ABBA nodding pattern twice, yielding a total on-source integration time of 40\,min. A standard A0-type star was observed immediately after the science observations with a similar airmass to ensure a proper correction for the telluric contribution. Reduction of the raw data was accomplished using IRAF routines. The AB nodded frames were subtracted to remove the background emission contribution. Individual sky-subtracted frames were then registered using the bright reference star and stacked together to produce one high signal-to-noise image. The spectra of J1551$-$2134 and the telluric standard star were optimally extracted and wavelength calibrated using a NeNeAr lamp exposure. After removal of the intrinsic features (typically hydrogen lines) of the A0-type star, the calibration spectrum was divided into the corresponding target data to remove telluric absorptions and instrumental spectral response. Finally, the data were multiplied by a black body curve of 9700\,K to restore the spectral slope of J1551$-$2134.

\begin{table}
\caption{VIMOS observing log, completeness and limiting magnitudes. \label{u:log}}
\centering
\scriptsize
\tabcolsep=0.09cm
\begin{tabular}{clcccccc}
\hline\hline
\noalign{\smallskip}
Pointing &~~~RA (J2000)~~DEC  & Filter &  Date & Texp & Seeing & Comp$.$ & Lim$.$  \\
&~~~($^{\rm h~~~m~~~s}$)  ~~~~($^{\rm o~~~'~~~''}$) & & & (s) & (\arcsec) & (mag) & (mag)  \\
\hline
\noalign{\smallskip}
1 &   15:53:30.0$-$21:19:59.9  & $I$  & 2009 Apr 02  &   3$\times$160 & 0.57 & 23.1 & 23.8 \\
   &                                                &  $z$ & 2009 Apr 02 &   3$\times$105 & 0.53 & 22.8 & 23.5\\
2 &   15:54:38.8$-$21:55:59.9  & $I$  & 2009 Apr 02 &   3$\times$160 & 0.66 & 22.8 & 23.8\\
   &                                                &  $z$ & 2009 Apr 02 &   3$\times$105 & 0.55 & 22.4 & 23.3\\
3 &   15:54:38.8$-$21:37:59.9  & $I$  & 2009 Apr 02 &   3$\times$160 & 0.64 & 22.7 & 23.6\\
   &                                                &  $z$ & 2009 Apr 02 &   3$\times$105 & 0.61 & 22.4 & 23.1\\
4 &   15:54:38.8$-$21:19:59.9  & $I$  & 2009 Apr 02 &   3$\times$160 & 0.58 & 23.0  & 23.9 \\
   &                                                &  $z$ & 2009 Apr 02 &   3$\times$105 & 0.58 & 23.0 & 23.7\\
5 &   15:47:46.0$-$21:55:59.9  & $I$  & 2009 Apr 20 &   3$\times$160 & 0.86 & 22.0 & 23.0\\
   &                                                &  $z$ & 2009 Apr 20 &   3$\times$105 & 0.78 & 21.7 & 22.6\\
6 &   15:47:46.0$-$21:37:59.9  & $I$  & 2009 Apr 20 &   3$\times$160 & 0.73 & 22.5 & 23.3\\
   &                                                &  $z$ & 2009 Apr 20 &   3$\times$105 & 0.63 & 22.1 & 23.1\\
7 &   15:47:46.0$-$21:19:59.9  & $I$  & 2009 Apr 20 &   3$\times$160 & 0.79 &  22.3 & 23.2\\ 
   &                                                &  $z$ & 2009 Apr 20 &   3$\times$105 & 0.74 & 22.0 & 22.7\\
8 &   15:48:54.7$-$21:55:59.9  & $I$  & 2009 Apr 20 &   3$\times$160 & 0.66 & 22.4 & 23.7\\
   &                                                &  $z$ & 2009 Apr 20 &   3$\times$105 & 0.56 & 22.3 & 23.2\\
9 &   15:48:54.7$-$21:37:59.9  & $I$  & 2009 Apr 20 &   3$\times$160 & 0.63 & 22.0 &23.3 \\ 
   &                                                &  $z$ & 2009 Apr 20 &   3$\times$105 & 0.66 & 21.9 &22.6 \\
10 & 15:48:54.7$-$21:19:59.9  & $I$  & 2009 Apr 20 &   3$\times$160 & 0.66 & 22.6 & 23.5\\
   &                                                &  $z$ & 2009 Apr 20 &   3$\times$105 & 0.66 & 22.2 & 23.0\\
11 & 15:51:12.4$ $-21:37:59.9 & $I$  & 2009 Apr 21 &   3$\times$160 & 0.53 & 23.3 & 24.1\\
   &                                                &  $z$ & 2009 Apr 21 &   3$\times$105 & 0.50 & 22.8 & 23.8\\
12 & 15:51:12.4$ $-21:19:59.9 & $I$  & 2009 Apr 21 &   3$\times$160 & 0.52 & 23.2 & 24.1\\
   &                                                &  $z$ & 2009 Apr 21 &   3$\times$105 & 0.55 & 22.9 & 23.5\\
13 & 15:52:21.1$-$21:37:59.9 & $I$  &  2009 Apr 21 &   3$\times$160 & 0.68 & 22.8 & 23.6\\
   &                                                &  $z$ & 2009 Apr 21 &   3$\times$105 & 0.59 & 22.4 & 23.2\\
14 & 15:52:21.1$-$21:55:59.9 & $I$  &  2009 Apr 22 &   3$\times$160 & 0.65 & 23.1 & 23.9\\
   &                                                &  $z$ & 2009 Apr 22 &   3$\times$105 & 0.60 & 22.6 & 23.3\\
15 & 15:52:21.1$-$21:19:59.9 & $I$  &  2009 Apr 22 &   3$\times$160 & 0.52 & 23.2 & 24.0\\
   &                                                &  $z$ & 2009 Apr 22 &   3$\times$105 & 0.54 & 22.7 & 23.4\\
16 & 15:53:30.0$-$21:55:59.9 & $I$  &  2009 Apr 22 &   3$\times$160 & 0.54 & 23.1 & 24.0\\
   &                                                &  $z$ & 2009 Apr 22 &   3$\times$105 & 0.48 & 22.9 & 23.4\\
17 & 15:53:30.0$-$21:37:59.9 & $I$  &  2009 Apr 22 &   3$\times$160 & 0.53 & 23.0 & 23.9\\
   &                                                &  $z$ & 2009 Apr 22 &   3$\times$105 & 0.49 & 22.8 & 23.4\\
18 & 15:50:03.6$-$21:55:59.9 & $I$  &  2009 Apr 22 &   3$\times$160 & 0.71 & 22.7 & 23.6\\
   &                                                &  $z$ & 2009 Apr 22 &   3$\times$105 & 0.66 & 22.2 & 22.9\\
19 & 15:50:03.6$-$21:37:59.9 & $I$  &  2009 Apr 22 &   3$\times$160 & 0.64 & 22.8 & 23.7\\
   &                                                &  $z$ & 2009 Apr 22 &   3$\times$105 & 0.54 & 22.6 & 23.1\\
20 & 15:50:03.6$-$21:19:59.9 & $I$  &  2009 Apr 22 &   3$\times$160 & 0.55 & 22.8 & 23.7\\
   &                                                &  $z$ & 2009 Apr 22 &   3$\times$105 & 0.60 & 22.3 & 23.0\\
21 & 15:51:12.4$-$21:55:59.9 & $I$  &  2009 Apr 22 &   3$\times$160 & 0.60 & 22.6 & 23.5\\
    &                                               &  $z$ & 2009 Apr 22 &   3$\times$105 & 0.56 & 22.4 & 23.0\\
\hline
\end{tabular}
\end{table}

%%%%%%%%%%%%%%%%%%%%%%%%%%%%%%%%%%

\section{Selection of USco member candidates}\label{selec}
\subsection{VIMOS--UKIDSS $zJ$ search}\label{cri1}
Given the depth of the entire photometric datasets (VIMOS and UKIDSS) available to us and the fact that we were searching for faint sources with red optical to near-infrared colors, we concluded that the combined VIMOS $z$- and UKIDSS $J$-band survey may yield promising USco member candidates. We retrieved UKIDSS DR8 $J$-band photometry for the sources identified in our VIMOS data using a correlation radius of 2\farcs0. Sometimes, two VIMOS sources that are close to each other and have very different brightness were cross-correlated with one single, typically bright UKIDSS source (wrongly identified as a probable galaxy by the UKIDSS algorithms in many cases), possibly because of the differing spatial scale and depth of the two catalogs. This resulted in false red $z-J$ colors that may contaminate the $zJ$ color-magnitude diagram. To prevent this from happening, we discarded those VIMOS faint objects with bright sources at less than 2\farcs0. Nevertheless, we first secured that none of them has red VIMOS $I-z$ colors in the region of interest as defined in Section~\ref{seciz}. 

\begin{figure}
\centering
\includegraphics[width=0.45\textwidth]{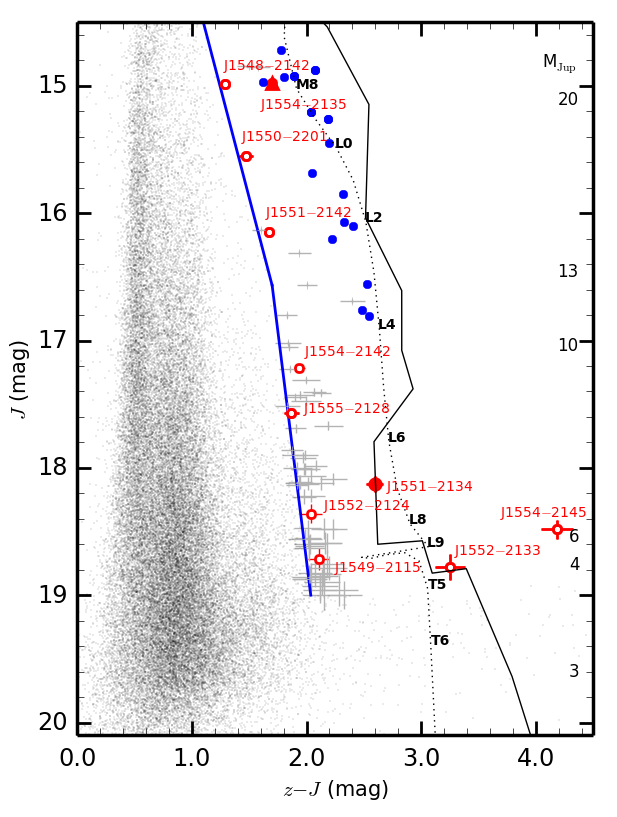}\\
\caption{Color-magnitude diagram, $J$ versus $z-J$, of the VIMOS--UKIDSS search. Previously known USco members are shown with blue filled dots. The separator (two blue straight lines) used for the selection of USco member candidates goes from $J$\,=\,14.5 to 19.0\,mag. The 11 unresolved sources --- labeled with their abridged names --- that pass the $zJ$ criterion are plotted as red symbols: the new candidate is shown with a red solid dot, and the previously known USco member recovered here, USco\,J155419.99$-$213543.1 \citep{lodieu06,lodieu08}, is depicted with a red filled triangle. The resolved sources complying with our $zJ$ criterion are plotted as gray crosses, and the bulk of field detections is shown with small black dots. For comparison purposes, the average sequence of field M, L, and T dwarfs normalized to the mid-M members of USco (not shown) is over-plotted with a dotted line. The 5-Myr isochrone of \citet{chabrier00model} is included with a solid line (we adopted a distance of 145 pc). Its corresponding masses (Jovian units) are indicated on the right side.  %The completeness (10$\sigma$) and the limit detection (4$\sigma$) are indicated with dashed and solid lines, respectively. 
\label{ujzj}}
\end{figure}

We built the $zJ$ color-magnitude diagram depicted in Figure~\ref{ujzj}. The dynamic range of the VIMOS--UKIDSS survey covers 5.5 mag from $J$\,=\,14.5 through 20.0\,mag. The faint end is imposed by the UKIDSS limiting magnitude and the bright limit is such that avoids the highly saturated sources and sources with magnitudes in the non-linear regime of the detectors. The completeness of the $zJ$ survey is given by the interval $J$\,=\,14.5--19.0\,mag. As seen in Figure~\ref{ujzj}, the USco photometric sequence is well delineated by known members of the association (plotted as filled blue dots). In the surveyed area and with magnitudes fainter than $J$\,=\,14.5\,mag, there are seven USco candidates previously published in the literature. They were detected photometrically by \citet{lodieu06,lodieu07} and all have spectroscopic follow-up observations in \citet{lodieu08,lodieu11}. These objects show colors, proper motions, H$\alpha$ emission, and other spectroscopic features typical of low-gravity atmospheres, which support their membership in the USco association.  

The separator that we used to discriminate USco member candidates from field sources in Figure~\ref{ujzj} is formed by two straight lines. One of these lines covers the interval $J$\,=\,14.5--16.5\,mag and runs parallel to the USco sequence of known objects shifted by $\sim$4-$\sigma$ towards blue $z-J$ colors, where $\sigma$ represents the color dispersion of the USco photometric sequence. The other line separator was employed for the interval $J$\,=\,16.5--19.0\,mag and was based on the separator proposed by \citet{lodieu07}: it goes from ($z -J$, $J$) = (1.7, 16.5) to (2.0, 19.0) mag. The defined field--USco separator is shown with a solid line in Figure~\ref{ujzj}. USco member candidates must fall to the red of the separator line. This is a rather conservative selection criterion intended to identify all possible candidates. We acknowledged that additional criteria were required to clean the list of photometric candidates and avoid contaminants. 

Within the VIMOS--UKIDSS survey completeness ($J = 14.5-19$\,mag), a total of 92 sources were found populating the red side of the defined separator in Figure~\ref{ujzj}. Of them, 65 are resolved (their FWHMs are at least 1.5 times higher than the width provided by the average seeing), and 27 sources appear to be point-like objects. We did not consider the 65 resolved objects in our list of USco candidates (see next). We checked the catalogued aperture UKIDSS photometry by obtaining the PSF photometry of the 27 unresolved objects, the majority of which are quite faint and close to the completeness magnitude of the survey. The new PSF photometry located 16 out of 27 sources to the blue of our $zJ$ selection criterion (many of these objects have aperture photometry contaminated by nearby bright stars), and they were rejected. Only 11 unresolved sources remained as $zJ$ USco member candidates. They are depicted with red symbols and are labelled with their abridged names in Figure~\ref{ujzj}. In Table~\ref{los12}, we provide their photometry. With the exception of three objects, most lie rather close to the artificial boundary defined to separate USco candidates from field sources.                                                         

\begin{table*}[htdp]
\caption{USco member candidates selected from the combined VIMOS--UKIDSS $zJ$ search.\label{los12}}
\centering
\scriptsize
\tabcolsep=0.045cm
\begin{tabular}{clccccccccc}
\hline\hline
\noalign{\smallskip}
~~~RA (J2000)~~DEC  & \multicolumn{1}{c}{Name} & $I$ & $z$ & $Z$ & $Y $& $J$ & $H$ & $K$ & $W1$ & $W2$\\
~~~($^{\rm h~~~m~~~s}$)  ~~~~($^{\rm o~~~'~~~''}$) &&(mag)&(mag)&(mag)&(mag)&(mag)&(mag)&(mag)&(mag)&(mag)\\
\hline
\noalign{\smallskip}
15 54 19.9$-$21 35 43 &USco\,J155419.9$-$213543\tablefootmark{a} &  17.680$\pm$0.113    &16.670$\pm$0.026  & 16.852$\pm$0.009  &  15.847$\pm$0.005 & 14.975$\pm$0.004 & 14.321$\pm$0.003 &   13.746$\pm$0.003 & 13.260$\pm$0.025 & 12.584$\pm$0.027 \\%032_655  
15 48 00.9$-$21 42 42 &USco\,J154800.9$-$214242                  &  16.918$\pm$0.110    &16.278$\pm$0.041  & 16.276$\pm$0.007  &  15.739$\pm$0.005 & 14.987$\pm$0.004 & 14.336$\pm$0.003 &   13.960$\pm$0.004 & 13.775$\pm$0.027 & 13.586$\pm$0.041 \\%064_692  
15 50 11.5$-$22 01 22 &USco\,J155011.5$-$220122\tablefootmark{b} &  18.044$\pm$0.071    &17.018$\pm$0.064  & 17.209$\pm$0.012  &  16.337$\pm$0.008 & 15.548$\pm$0.006 & 14.979$\pm$0.006 &   14.572$\pm$0.007 & 14.397$\pm$0.030 & 14.193$\pm$0.053 \\%204_912  
15 51 32.7$-$21 42 43 &USco\,J155132.7$-$214243                  &  18.600$\pm$0.104    &17.823$\pm$0.034  & 17.951$\pm$0.018  &  17.000$\pm$0.011 & 16.148$\pm$0.010 & 15.593$\pm$0.009 &   15.108$\pm$0.011 & 14.923$\pm$0.039 & 14.776$\pm$0.091 \\%114_2135
15 54 50.8$-$21 42 32 &USco\,J155450.8$-$214232                  &  20.313$\pm$0.122    &19.153$\pm$0.045  & 19.409$\pm$0.048  &  18.256$\pm$0.027 & 17.214$\pm$0.024 & 16.506$\pm$0.022 &   15.966$\pm$0.025 & 15.790$\pm$0.059 & 15.669$\pm$0.173 \\%034_1590
15 55 08.4$-$21 28 42 &USco\,J155508.4$-$212842                  &  20.439$\pm$0.107    &19.435$\pm$0.060  & 19.503$\pm$0.060  &  18.481$\pm$0.028 & 17.566$\pm$0.030 & 16.964$\pm$0.030 &   16.388$\pm$0.035 & \nodata          & \nodata          \\%044_111  
15 51 50.2$-$21 34 57 &USco\,J155150.2$-$213457\tablefootmark{c} &  21.896$\pm$0.094    &20.724$\pm$0.043  & 20.658$\pm$0.163  &  19.608$\pm$0.090 & 18.127$\pm$0.060 & 17.219$\pm$0.042 &   16.410$\pm$0.039 & 15.845$\pm$0.063 & 15.261$\pm$0.123 \\%132_794  
15 52 34.2$-$21 24 47 &USco\,J155234.2$-$212447\tablefootmark{d} &  21.565$\pm$0.122    &20.399$\pm$0.054  & \nodata           &  19.556$\pm$0.092 & 18.359$\pm$0.073 & 17.654$\pm$0.065 &   17.059$\pm$0.075 & 16.740$\pm$0.110 & 16.233$\pm$0.266 \\%154_910  
15 49 54.5$-$21 15 04 &USco\,J154954.5$-$211504\tablefootmark{d} &  22.070$\pm$0.080    &20.821$\pm$0.042  & \nodata           &  19.935$\pm$0.087 & 18.715$\pm$0.083 & 17.836$\pm$0.070 &   17.278$\pm$0.077 & 17.172$\pm$0.175 & 15.864$\pm$0.207 \\%222_1419
\hline
\noalign{\smallskip}
%15 52 12.0$-$21 25 47&USco\,J155212.0$-$212547                  &  20.643$\pm$0.103    &20.188$\pm$0.098  & 19.270$\pm$0.050  &  18.909$\pm$0.049 & 18.180$\pm$0.063 & 17.430$\pm$0.051 &   16.658$\pm$0.049 & 15.544$\pm$0.052 & 14.944$\pm$0.097 \\%153_936  
15 52 10.4$-$21 33 41 &USco\,J155210.4$-$213341                  &  22.485$\pm$0.098    &22.027$\pm$0.084  & \nodata           &  19.623$\pm$0.090 & 18.778$\pm$0.102 & 18.853$\pm$0.186 &   $\geq$18.7       & \nodata          & \nodata          \\%132_1314
15 54 54.1$-$21 45 55 &USco\,J155454.1$-$214555\tablefootmark{e} &  23.296$\pm$0.140    &22.669$\pm$0.118  & 19.762$\pm$0.064  &  19.315$\pm$0.066 & 18.483$\pm$0.075 & 18.203$\pm$0.105 &   18.380$\pm$0.231 & \nodata          & \nodata          \\%034_410  
\hline
\end{tabular}
\tablefoot{Right ascension (RA) and declination (DEC) coordinates as derived from the VIMOS astrometric calibration. Top and bottom panels list the 11 photometric USco candidates from the combined VIMOS--UKIDSS survey (Section~\ref{cri1}). The top panel indicates those candidates that were also found in the VIMOS-only search (Section~\ref{seciz}). \tablefoottext{a}{USco member discovered and confirmed by \citet{lodieu06,lodieu08}. It has $W3$\,=\,11.034$\pm$0.136\,mag.}
\tablefoottext{b}{High proper motion source (Section~\ref{pm}) discovered by \citet{deacon09}.}
\tablefoottext{c}{Candidate with spectroscopic follow-up (Section~\ref{anal_spectrum}).}
\tablefoottext{d}{Non-photometric candidate according to the diagrams of Figure\,\ref{udemas}.}
\tablefoottext{e}{We provided $J$-band PSF photometry. $ZYHK$ photometry as in the UKIDSS catalog (we checked the PSF photometry of these bands agrees with UKIDSS aperture photometry within the quoted error bars). }
}
 \end{table*}%

Of the 11 USco candidates, two were previously known: USco J155419.99$-$213543.1 (J1554$-$2135) and USco\,J155011.5$-$220122 (2MASS\,J15501151$-$2201213). The former object was first identified from an UKIDSS-only survey by \citet{lodieu06}. In \citet{lodieu08} it was confirmed as an M8.0 bonafide member of the USco association. The latter object was first reported to have a high proper motion from the cross-correlation of UKIDSS and 2MASS catalogs by \citet{deacon09}. See also Section~\ref{pm}. Except for these two sources, none of the USco candidates published by other groups has a magnitude brightness within the dynamic range or lies within the effective area of our study. 
%%%%%%%%%%%%%%%%%%%%%%%%%%%%%%%%%%

\subsection{VIMOS $Iz$ selection }\label{seciz}
\begin{figure}[pt!]
\centering
\includegraphics[width=0.45\textwidth]{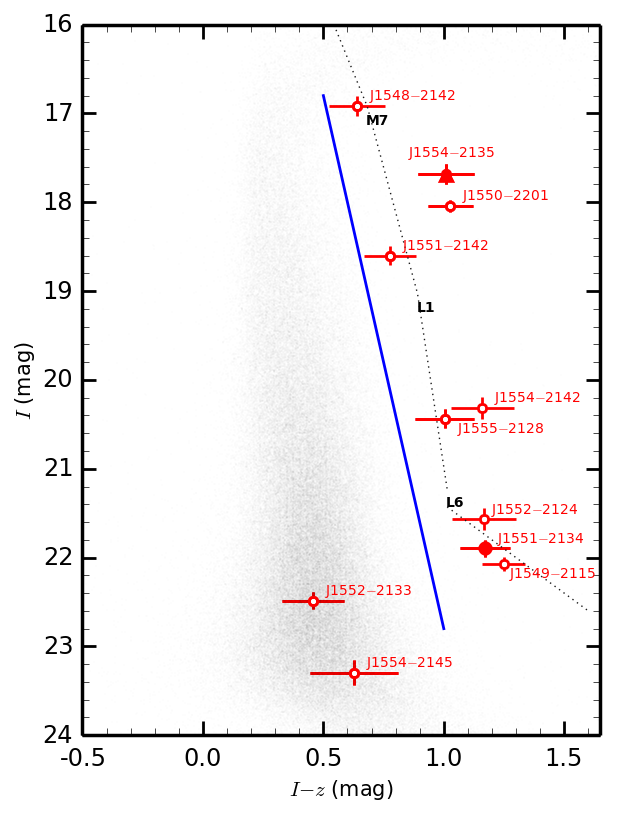}\\
\caption{Color-magnitude diagram, $I$ versus $I-z$, of the VIMOS-only search. Symbols as in Figure~\ref{ujzj}. The solid blue line separates USco member candidates from other sources. The VIMOS--UKIDSS candidates J1552$-$2133 and J1554$-$2145 fail the $Iz$ criterion for USco candidates.\label{uiiz}}
\end{figure}

To provide a more reliable list of USco member candidates, we combined the previous $zJ$ search with a selection based on VIMOS data only. We used the $I$-band versus $I-z$ color-magnitude diagram depicted in Figure~\ref{uiiz}. Our criterion to select USco member candidates followed the prescription proposed by \citet{ardila00} for a similar diagram. These authors dealt with candidates in the magnitude interval $I$\,=\,13--19\,mag, which includes the late-M types. Our survey is deeper than theirs; we thus linearly extrapolated the proposed criterion towards fainter magnitudes and the L types as shown by the solid line in Figure~\ref{uiiz}. This extrapolation falls to the blue of the photometric sequence of field L dwarfs down to the completeness of the VIMOS data. The field sequence displayed in Figures~\ref{ujzj} and~\ref{uiiz} was built by using data from \citet{hewett06}, \citet{patten06}, and \citet{leggett07} as described in Section~\ref{others}.

Nine candidates from the VIMOS--UKIDSS $zJ$ search were found to fulfil the $Iz$ criterion, i.e., they lie to the red of the field--cluster photometric separator. The nine sources are included in the top panel of Table~\ref{los12}. Only two $zJ$ candidates failed the $Iz$ selection process (bottom panel of Table~\ref{los12}) because they display blue $I-z$ colors despite having the faintest $I$ magnitudes in our list of candidates. Surprisingly, they were the two reddest candidates in the $zJ$ search. Next, we investigated these two sources in detail.

In order to confirm the photometry of J1554$-$2145, we collected $J$-band images with the Long-slit Intermediate Resolution Infrared Spectrograph (LIRIS; \citet{manchado04}) on the William Herschel Telescope (WHT) on 2012 June 15. LIRIS has a HAWAII detector of 1024\,$\times$\,1024\,pix$^2$ with a plate scale of 0$\farcs$25 per pixel. In imaging mode, LIRIS has a field of view of 4.27\,$\times$\,4.27\,arcmin$^2$. The LIRIS image provided us with a time baseline of 7\,yr with respect to the older UKIDSS data, and of 5.2\,yr with respect to the VIMOS images. The LIRIS $J$-band data were acquired with a dithering pattern of nine positions over the detector; individual exposure time was 50\,s per dither, and the total exposure time was 2250\,s. Observing conditions were photometric with a seeing of 1.4\arcsec. Data reduction included flat field correction and sky subtraction using routines within the IRAF environment. Individual images were aligned and combined to produce deep data in the $J$ band. The final LIRIS image has a limiting magnitude of $J$\,=\,20.7\,mag at the 4-$\sigma$ level. This photon depth would have been sufficient to guarantee the detection of J1554$-$2145 with an excellent signal-to-noise ratio (S/N) if the object's brightness were that of the UKIDSS catalog ($J = 18.483$ mag). However, this particular source was not detected in the LIRIS image, indicating that at the time of the LIRIS observations, J1554$-$2145 had $J \ge 20.7$\,mag, in highly contrast with the UKIDSS $J$ band measurement. Based on this result, we concluded that J1554$-$2145 is a variable source whose nature cannot be unveiled with our current data. It might actually be an USco member; there are known young sources with strong photometric variations due to circumbinary disks in Orion, or stars like KH\,15D \citep{johnson04} and CHS\,7797 \citep{rodriguez12}. To the best of our knowledge, J1554$-$2145 has not been previously identified in any variable or extragalactic source catalog or in supernovae databases.

The VIMOS $z$ and UKIDSS $J$ equatorial coordinates of J1552$-$2133 differ by 1\farcs5. This suggests that either we identified two different sources at optical and near-infrared wavelengths or the source has a high proper motion of $\approx$0.4\,arcsec\,yr$^{-1}$. We were convinced that the VIMOS $z$ and $I$ identifications correspond to one source because both data were taken at the same time and both $z$ and $I$ coordinates agree within the astrometric error bars. We thus relied on the $I-z$ color. Unfortunately, we cannot confirm whether the UKIDSS detection corresponds to the VIMOS object. In any scenario, this source does not appear to be a member of USco according to our selection criteria: either its $I-z$ color is too blue or it has a high proper motion inconsistent with the stellar association. From now on, neither J1554$-$2145 nor J1552$-$2133 (bottom panel of Table~\ref{los12}) are considered as USco photometric candidates. Therefore, we were left with 9 candidates (two of which are known in the literature, that successfully pass the $zJ$ and $Iz$ photometric criteria down to the completeness magnitude of the VIMOS--UKIDSS survey.

%J1552$-$2125 shows a $I-z$ color bluer than the one expected for truly USco members of similar magnitudes although its $z-J$ color is the reddest from all the twelve original $zJ$ candidates. Its VIMOS $z$-band photometry differs significantly from its UKIDSS cataloged $z$ band magnitude. Its further location in the different diagrams of Section\,\ref{others} confirms that this source is likely a galaxy and therefore is discarded as a reliable candidate.

To fully exploit the combined VIMOS--UKIDSS survey between completeness and limiting magnitudes, we searched for sources with the following photometric criteria, which are valid to identify cool dwarfs with spectral types later than early-L in USco: $J>19.0$, $I>22.0$, $z-J>2$, and $I-z>1$ mag, or $I$-band non-detections. Ten objects were found with colors redder than the $zJ$ field--USco boundary in Figure~\ref{ujzj}, seven of which turned out to be false detections close to the spikes of very bright stars. One showed an extended profile with 3.2 times the stellar FWHM, and the remaining two unresolved sources lied near bright stars. We carried out their PSF photometry using the UKIDSS images and obtained that their new colors were not compatible with our criteria (they moved to the cloud of field objects, i.e., to the blue side of the separator in Figure~\ref{ujzj}). In short, the search for new USco member candidates with VIMOS and UKIDSS magnitudes beyond completeness yielded no new objects of interest. Therefore, our USco candidates are those listed in the top panel of Table~\ref{los12}.

%%%%%%%%%%%%%%%%%%%%%%%%%%%%%%%%%%

\subsection{Additional photometric criteria}\label{others}
\begin{figure*}
\centering
\includegraphics[width=0.45\textwidth]{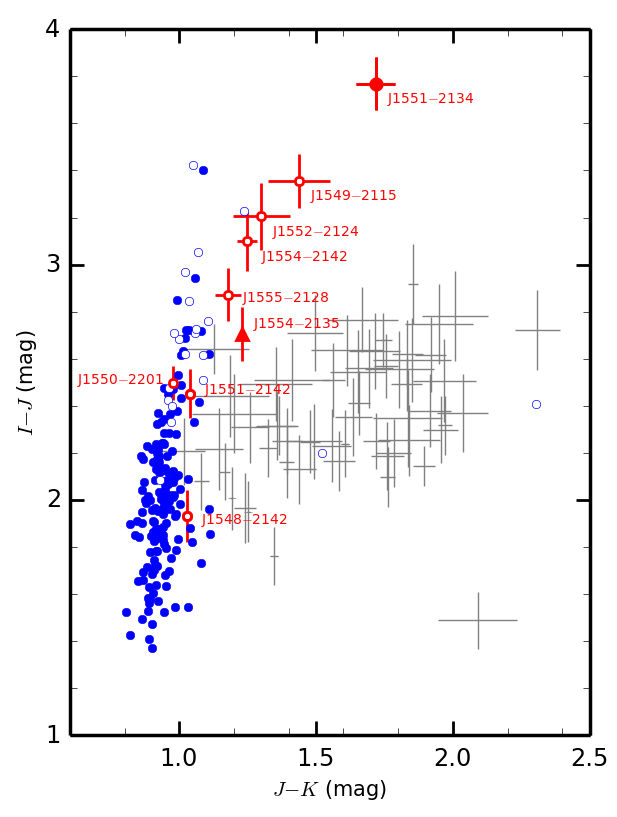}
\includegraphics[width=0.45\textwidth]{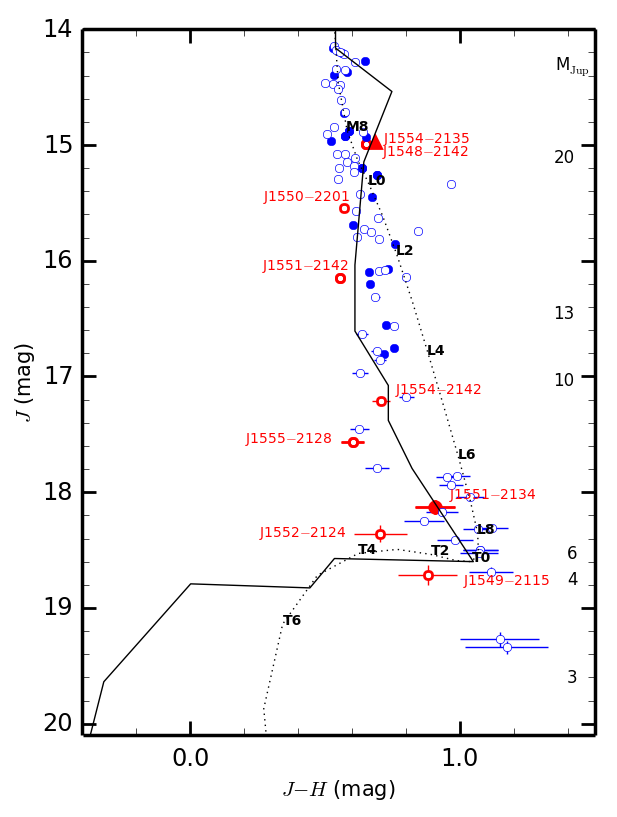}\\
\includegraphics[width=0.45\textwidth]{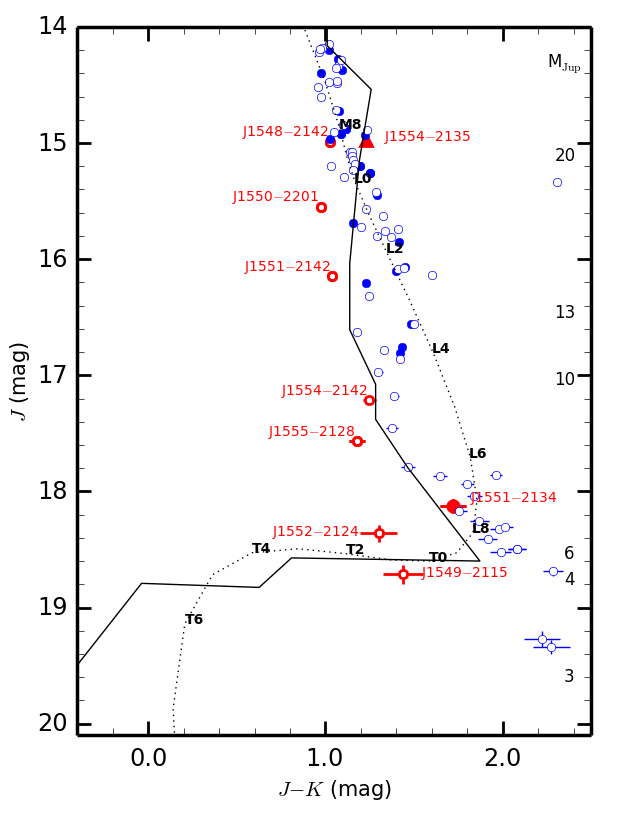}
\includegraphics[width=0.45\textwidth]{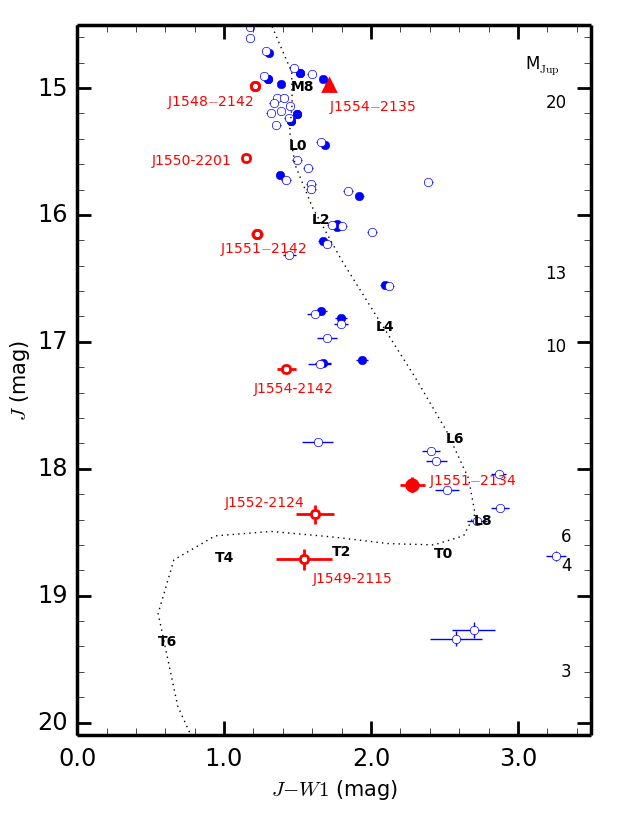}
\caption{Color--color (top left panel) and color--magnitude diagrams (top right and bottom panels) combining VIMOS, UKDISS, and {\sl WISE} photometry. The nine $zJ$ USco candidates are plotted as red symbols and are labelled. Confirmed USco members from \citet{ardila00} and \citet{lodieu11} are shown with blue solid circles, and USco candidates from \citet{lodieu13surv} are illustrated with blue open circles. Other symbols as in Figure~\ref{ujzj}. 
\label{udemas}}
\end{figure*}

To provide further robustness to our list of USco member candidates, we built additional color--magnitude and color--color diagrams. Figure~\ref{udemas} depicts various of these diagramas using the VIMOS, UKIDSS and {\sl WISE} photometry, thus covering the wavelength interval between 0.8 and 3.4 $\mu$m. The nine candidates from the top panel of Table~\ref{los12} are indicated with red symbols and are labelled in all panels of the Figure. We also included the known USco confirmed members and photometric candidates published by \citet{ardila00}, \citet{lodieu11}, and \citet{lodieu13surv}, 

The $I-J$ versus $J-K$ color--color panel of Figure~\ref{udemas} is useful to discriminate extragalactic sources from very red dwarfs (see the discussion by \citealt{bihain09}). In this panel, we added those resolved objects found in Section~\ref{cri1} whose FWHMs were 1.5 times higher than the stellar PSF. All of these objects occupy a particular region of the color--color diagram: they tend to have red $J-K$ values while their $I-J$ colors do not exceed the 3 mag boundary. These colors, where the $K$ band is particularly red, are typical of galaxies \citep{franx03}. None of the nine USco photometric candidates fall within this region; on the contrary, they follow, and actually extrapolate, the color-color sequence defined by the previously confirmed USco members. This adds consistency to our photometric selection of unresolved objects described in previous Sections. 

The USco candidates J1552$-$2124 and J1549$-$2115, which have the faintest $J$-band magnitudes in Table~\ref{los12}, display $J-H$, $J-K$, and $J-W1$ colors bluer than other USco objects. As shown in Figure~\ref{udemas} and from comparison with the field sequence, these two objects might have early-T spectral types if they were USco members and if the USco sequence were described by the field sequence of red dwarfs. The average location of field dwarfs between the mid-M types and the late-T types shown in Figures~\ref{ujzj}--\ref{udemas} was constructed by calculating the mean magnitudes and colors for each spectral type \citep{hewett06} and by using the equations published by \citet{dupuy12}. It was normalized to the location of thirteen USco M5.5-type dwarfs taken from \citet{slesnick08}. However, J1552$-$2124 and J1549$-$2115 do not display $I-z$ and $z-J$ colors fully compatible with T0--T5 dwarfs and their blue deviation from the trend delineated by other USco candidates in the color--magnitude diagrams of Figure~\ref{udemas} was interpreted as a likely non-membership signature.

From the combined 1.17-deg$^2$ VIMOS--UKIDSS USco survey exploring the magnitude interval $J = 14.5-19$ mag, we identified a total of 7 photometric member candidates with colors covering the broad 0.8--3.4\,$\mu$m wavelength range consistent with membership in the young stellar association. They likely have spectral types between late-M and mid-L as it is inferred from the direct comparison of their colors with the indices of field, high gravity dwarfs.

 %%%%%%%%%%%%%%%%%%%%%%%%%%%%%%%%%%%% 

\subsection{Contamination}\label{ucont}
Our photometric search may suffer from some contamination, which we studied next. The main source of contaminants was expected to be due to Galactic field sources of M, L and T spectral types, and reddened galaxies. We derived the Galactic contamination contribution following the prescription given in \citet[and references therein]{caballero08contamination}. In the range of magnitudes $J$\,=\,14.5--19\,mag (the $zJ$ search), a total of $\sim$4.5 field M-, L-, and T-dwarfs are expected to pollute the explored regions of the color-magnitude diagrams of previous Sections. These contaminant objects are likely interlopers distributed in spectral types and magnitude intervals as follows: $\sim$2.7 field M dwarfs would appear in the range $J$\,=\,14.5--16.0\,mag , and $\sim$1.6 field L dwarfs would be contaminating the faintest magnitude bin $J$\,=\,16--19\,mag. As for the field T dwarfs, the expected pollution was small with roughly $\sim$0.2 T-type objects at the faintest magnitudes of $J$\,=\,18--19\,mag. The contamination due to background giant red stars of M spectral types was expected to be negligible since the USco stellar association is located at latitudes above the Galactic plane ($b$\,$\sim$\,$+20$\,deg). The derived contamination by field cool dwarfs with spectral types similar to those we were seeking in USco suggested that, out of the 7 photometric candidates, only 2--3 would remain as USco members. 

Regarding the extragalactic contamination, the good seeing of the VIMOS data and the small size of the VIMOS pixel allowed us to easily distinguish extended sources down to $J=19$ mag as described in Section~\ref{cri1}. The location of these objects in the $I-J$ versus $J-K$ color-color diagram of Figure\,\ref{udemas} diverges from that of USco members as discussed in Section\,\ref{others}. We checked the FWHM of our USco candidates in all the available UKIDSS images confirming that they were consistent with the stellar PSFs at various explored wavelengths. To estimate the number of unresolved red galaxies that may be contaminating our VIMOS--UKIDSS survey, we used the multicolor GOODS--MUSIC V2.0 catalog \citep{grazian06,santini09} in a similar manner as in \citet{bihain09}. In the magnitude range $J$\,=\,14.5--20.0\,mag, i.e., down to the UKIDSS $J$-band limiting magnitude, we searched for GOOD--MUSIC sources that comply with the photometric selection criteria described in Sections~\ref{cri1} and~\ref{seciz} and found none. Furthermore, a list of USco member candidates free of extragalactic unresolved red contaminants can be produced by performing a proper motion analysis of the candidates (see Section~\ref{pm}), since very distant sources do not show significant motion at all while USco has a distinctive proper motion.

%%%%%%%%%%%%%%%%%%%%%%%%%%%%%%%%%%%% 

\subsection{Proper motion analysis}\label{pm}
The proper motion of the young USco stellar association is $\mu_\alpha\,{\rm cos}\,\delta$\,=\,$-12.1$\,$\pm$\,1.6 and $\mu_\delta$\,=\,$-23.8$\,$\pm$\,1.9\,mas\,yr$^{-1}$ \citep{zacharias04}, which is measurable using our data and data from the UKIDSS archive. The final assessment of membership in USco is done by deriving the proper motions of the 7 photometric candidates.  We employed various combinations of optical and near-infrared images separated in time by several years for this purpose. One collection of images was formed by UKIDSS $J$ and $K$ data (2005 June 05) and VIMOS $I$ and $z$ images (this paper), which were taken $\approx$3.8 yr apart. Another collection of images comprised the UKIDSS $K$-band first and second epoch data (2005 June 05, 2011 March 16), providing a time baseline of $\approx$5.8 yr. For those photometric candidates fainter than $J$\,=\,18.0\,mag, we also used the UKIDSS $H$-band images (2005 June 05) to improve the S/N of the astrometric measurements. The UKIDSS and VIMOS data had an average seeing of 0\farcs8 and 0\farcs6, respectively. 

Proper motions were obtained from the comparison of the target coordinates (in pixels) with the positions of $\sim$20--30 unresolved sources within the area of $3\times3$\,arcmin$^2$, except for except for J1551$-$2134, for which a larger area of $4\times4$\,arcmin$^2$ was analyzed. Pixel transformations were derived employing third/fourth order polynomials and the GEOMAP routine within IRAF. The typical dispersion of the polynomial transformations was $\pm$0.1\,pix for the the right ascension and declination axis after rejecting reference sources that deviated by more than 2--3\,$\sigma$ from null motion, where $\sigma$ denoted the dispersion of the astrometric transformations. All of the unresolved reference sources defining the null motion were selected to have S/N higher than 15 in the flux peak with respect to the background noise. By considering the temporal difference between images, and the pixel scales and north--east orientation of the detectors, displacements in pixels were converted into the proper motion measurements provided in Table~\ref{pmlos9}. The quoted astrometric uncertainties were obtained by quadratically adding the dispersions of the polynomial transformations and the errors of the targets' centroids as provided by the automatic identification algorithms; the latter oscillated between $\pm$0.01 pix for the brightest photometric candidates and $\pm$0.3\,pix for the faintest candidates. The astrometric error bars of the faintest targets are clearly dominated by the UKIDSS centroid errors. Table~\ref{pmlos9} also contains the time baselines, the filters, and the S/N of the  targets as measured on the corresponding images.

\begin{table}[htdp!]
\caption{Proper motion measurements.\label{pmlos9}}
\centering
\scriptsize
\tabcolsep=0.1cm
\begin{tabular}{lcrcrr}
\hline\hline
\noalign{\smallskip}
Abriged &  $\Delta t$&\multicolumn{1}{c}{S/N\tablefootmark{a}}&Images &\multicolumn{1}{c}{$\mu_\alpha\cos\delta$} & \multicolumn{1}{c}{$\mu_\delta$}\\
name&(yr)&\multicolumn{1}{c}{1$^{\rm st}$/2$^{\rm nd}$}  &UKIDSS / VIMOS&\multicolumn{1}{c}{(mas\,yr$^{-1}$}) &\multicolumn{1}{c}{(mas\,yr$^{-1}$)}  \\
\hline
\noalign{\smallskip}
J1554$-$2135&3.82 &$>$15 / $>$15& $J$  / $z$  & $-$10.61\,$\pm$\,3.69 & $-$20.33\,$\pm$\,3.83   \\%032_655  
&3.82&$>$15 / $>$15 &   $K$  / $z$ & $-$14.01\,$\pm$\,2.89 & $-$21.76\,$\pm$\,3.72   \\%032_655  
&3.82&$>$15 / $>$15 &  $J$  / $I$  & $-$11.19\,$\pm$\,3.73 & $-$17.49\,$\pm$\,3.58   \\%032_655  
&3.82&$>$15 / $>$15 &  $K$  / $I$  & $-$17.97\,$\pm$\,2.22 & $-$13.88\,$\pm$\,2.26  \\%032_655 
&5.78&$>$15 / $>$15 &  $K$  / $K$  & $-$13.03\,$\pm$\,3.07 & $-$22.81\,$\pm$\,3.23  \\%032_655  
&&&Weighted value&$-$13.72\,$\pm$\,3.73&$-$20.50\,$\pm$\,3.83\\
\noalign{\smallskip}
\hline
\noalign{\smallskip}
J1548$-$2142 &   3.87  &$>$15 / $>$15& $J$  / $z$  &5.56\,$\pm$\,3.88 & $-$14.60\,$\pm$\,4.23  \\%064_692  
&3.87&$>$15 / $>$15&  $K$  / $z$ & 5.53\,$\pm$\,3.11 & $-$17.31\,$\pm$\,3.70   \\
&3.87&$>$15 / $>$15& $J$  / $I$  & 5.25\,$\pm$\,4.15 & $-$14.87\,$\pm$\,4.65   \\
&3.87&$>$15 / $>$15& $K$  / $I$  & 5.42\,$\pm$\,3.75 & $-$15.77\,$\pm$\,3.86   \\
&5.78&$>$15 / $>$15& $K$  / $K$  & 4.28\,$\pm$\,4.37 & $-$16.93\,$\pm$\,3.90   \\
&&&Weighted value&4.75\,$\pm$\,4.37&$-$16.48\,$\pm$\,4.65\\
\noalign{\smallskip}
\hline
\noalign{\smallskip}
J1550$-$2201\tablefootmark{b}&   3.88   &$>$15 / $>$15&$J$  / $z$  & 4.69\,$\pm$\,4.40 & $-$84.60\,$\pm$\,4.45   \\%204_912  
&3.88&$>$15 / $>$15&  $K$  / $z$ & $-$3.04\,$\pm$\,4.74 & $-$88.24\,$\pm$\,4.87  \\
&3.88&$>$15 / $>$15& $J$  / $I$ & 13.70\,$\pm$\,5.24 & $-$83.01\,$\pm$\,3.99    \\
&3.88&$>$15 / $>$15& $K$  / $I$ & 8.21\,$\pm$\,5.34 & $-$85.76\,$\pm$\,4.72  \\
&5.78&$>$15 / $>$15& $K$  / $K$ & 2.76\,$\pm$\,3.49 & $-$85.27\,$\pm$\,3.29  \\
&&&Weighted value& 3.79\,$\pm$\,5.34& $-$85.21\,$\pm$\,4.87\\
\noalign{\smallskip}
\hline
\noalign{\smallskip}
J1551$-$2142 &   3.88  &$>$15 / $>$15&$J$  / $z$  & 4.81\,$\pm$\,5.42 & $-$22.79\,$\pm$\,5.35 \\%114_2135
&3.88&$>$15 / $>$15& $K$  / $z$  & 8.67\,$\pm$\,4.21 & $-$22.21\,$\pm$\,4.18   \\
&3.88&$>$15 / $>$15&$J$  / $I$   & 5.76\,$\pm$\,5.27 & $-$17.51\,$\pm$\,5.09 \\
&3.88&$>$15 / $>$15&$K$  / $I$   & 11.95\,$\pm$\,3.86 & $-$20.41\,$\pm$\,4.06  \\
&5.78&$>$15 / $>$15&$K$  / $K$   & 7.15\,$\pm$\,3.69 & $-$27.46\,$\pm$\,3.37  \\
&&&Weighted value& 7.73\,$\pm$\,5.42 & $-$24.78\,$\pm$\,5.35\\
\noalign{\smallskip}
\hline
\noalign{\smallskip}
J1554$-$2142 &   3.82  &$>$15 / $>$15&$J$  / $z$  & $-$17.30\,$\pm$\,6.37 & 10.50\,$\pm$\,6.07    \\%034_1590
&3.82&$>$15 / $>$15& $K$  / $z$ & $-$21.07\,$\pm$\,5.71 & 9.06\,$\pm$\,5.03    \\
&3.82&$>$15 / $>$15& $J$  / $I$  & $-$18.27\,$\pm$\,6.54 & 9.48\,$\pm$\,6.47   \\
&3.82&$>$15 / $>$15& $K$  / $I$  & $-$21.90\,$\pm$\,5.76 & 11.25\,$\pm$\,4.75   \\
&5.78&$>$15 / $>$15& $K$  / $K$  & $-$23.97\,$\pm$\,5.10 & 10.81\,$\pm$\,4.43   \\
&&&Weighted value& $-$22.33\,$\pm$\,6.54 &  10.54\,$\pm$\,6.47 \\
\noalign{\smallskip}
\hline
\noalign{\smallskip}
J1555$-$2128 &   3.82 &14.7 / $>$15&$J$  / $z$  & $-$30.17\,$\pm$\,7.32 & $-$2.12\,$\pm$\,7.02 \\%044_111  
&3.82&$>$15 / $>$15& $K$  / $z$ & $-$21.60\,$\pm$\,4.50 & 4.27\,$\pm$\,5.34  \\
&3.82&14.7 / $>$15& $J$  / $I$  & $-$40.73\,$\pm$\,7.13 & 8.07\,$\pm$\,7.01  \\
&3.82&$>$15 / $>$15& $K$  / $I$  & $-$20.90\,$\pm$\,4.52 & $-$3.93\,$\pm$\,5.20  \\
&5.78&$>$15 / $>$15& $K$  / $K$  & $-$30.42\,$\pm$\,4.92 & $-$1.25\,$\pm$\,4.89  \\
&&&Weighted value&$-$28.37\,$\pm$\,7.32 & $-$0.31\,$\pm$\,7.02 \\
\noalign{\smallskip}
\hline
\noalign{\smallskip}
J1551$-$2134 &  3.88  &7.1 / $>$15&$J$ / $z$& 1.66\,$\pm$\,10.82 & $-$26.60\,$\pm$\,11.15  \\%132_794  
&3.88&10.5 / $>$15&$H$  / $z$  & $-$15.07\,$\pm$\,7.07 & $-$9.21\,$\pm$\,6.37   \\
&3.88&14.5 / $>$15& $K$ / $z$ & $-$8.73\,$\pm$\,4.68 & $-$13.95\,$\pm$\,4.74   \\
&3.88&7.1 / $>$15&$J$  / $I$   & $-$3.22\,$\pm$\,10.90 & $-$27.97\,$\pm$\,11.24   \\
&3.88&10.5 / $>$15& $H$  / $I$   & $-$19.88\,$\pm$\,7.03 & $-$12.87\,$\pm$\,6.55   \\
&3.88&14.5 / $>$15& $K$  / $I$   & $-$10.84\,$\pm$\,4.27 & $-$16.02\,$\pm$\,4.37   \\
&5.78&14.6 / $>$15& $K$  / $K$   & $-$5.68\,$\pm$\,4.85 & $-$22.85\,$\pm$\,5.00   \\
&&&Weighted value&$-$7.76\,$\pm$\,7.07&$-$19.72\,$\pm$\,6.55 \\
\noalign{\smallskip}
\hline
\noalign{\smallskip}
J1552$-$2124\tablefootmark{c} &   3.88   &6.0 / $>$15&$J$ / $z$  & $-$6.69\,$\pm$\,12.41&0.37\,$\pm$\,11.16     \\%154_910  
&3.88&7.9 / $>$15&  $H$ / $z$ & $-$3.78\,$\pm$\,6.97 & $-$23.15\,$\pm$\,6.68    \\
&3.88&8.0 / $>$15&  $K$  / $z$ & 9.02\,$\pm$\,6.10 & 8.93\,$\pm$\,7.04    \\
&3.88&6.0 / $>$15& $J$  / $I$   & $-$5.10\,$\pm$\,12.13 &$-$2.06\,$\pm$\,10.97  \\
&3.88&7.9 / $>$15& $H$  / $I$   & $-$5.22\,$\pm$\,6.71 & $-$23.73\,$\pm$\,6.63    \\
&3.88&8.0 / $>$15& $K$  / $I$   & 10.48\,$\pm$\,5.46 & 9.09\,$\pm$\,6.93    \\
&5.78&8.0 / $>$15& $K$  / $K$   & 5.23\,$\pm$\,5.38 & $-$0.88\,$\pm$\,6.14    \\
&&&Weighted value&4.24\,$\pm$\,10.90&$-$3.28\,$\pm$\,11.24 \\
\noalign{\smallskip}
\hline
\noalign{\smallskip}
J1549$-$2115\tablefootmark{c} &   3.88   &6.4 / $>$15& $J$  / $z$  &$-$20.01\,$\pm$\,12.11 & $-$12.04\,$\pm$\,10.70  \\%222_1419
&3.88&6.2 / $>$15&$H$  / $z$  & 12.09\,$\pm$\,9.43 & $-$9.29\,$\pm$\,9.50   \\
&3.88&8.3 / $>$15&$K$  / $z$  & $-$2.90\,$\pm$\,8.06 & $-$0.95\,$\pm$\,8.30   \\
&3.88&6.4 / $>$15&$J$  / $I$  & $-$24.98\,$\pm$\,11.81 & $-$14.31\,$\pm$\,10.75  \\
&3.88&6.2 / $>$15&$H$  / $I$   & 10.82\,$\pm$\,9.31 & $-$10.19\,$\pm$\,8.88   \\
&3.88&8.3 / $>$15&$K$  / $I$   & $-$4.75\,$\pm$\,7.96 & $-$7.39\,$\pm$\,7.69   \\
&5.78&7.6 / $8.1$&$K$  / $K$   & $-$0.11\,$\pm$\,6.22 & $-$3.00\,$\pm$\,6.26   \\
&&&Weighted value&$-$1.10\,$\pm$\,12.11& $-$5.11\,$\pm$\,10.75\\
\hline
\end{tabular}
\tablefoot{ 
\tablefoottext{a}{Signal-to-noise ratio of the photometric candidates on the corresponding UKIDSS and VIMOS images.}
\tablefoottext{b}{Known high proper motion source (2MASS\,J15501151$-$2201213). Our proper motion measurement agrees within 1-$\sigma$ with the value reported by \citet{deacon09}}
\tablefoottext{c}{Candidates discarded in Section\ref{others}.} 
}
\end{table}%

\begin{figure}[ht!]
\centering
\includegraphics[width=0.5\textwidth]{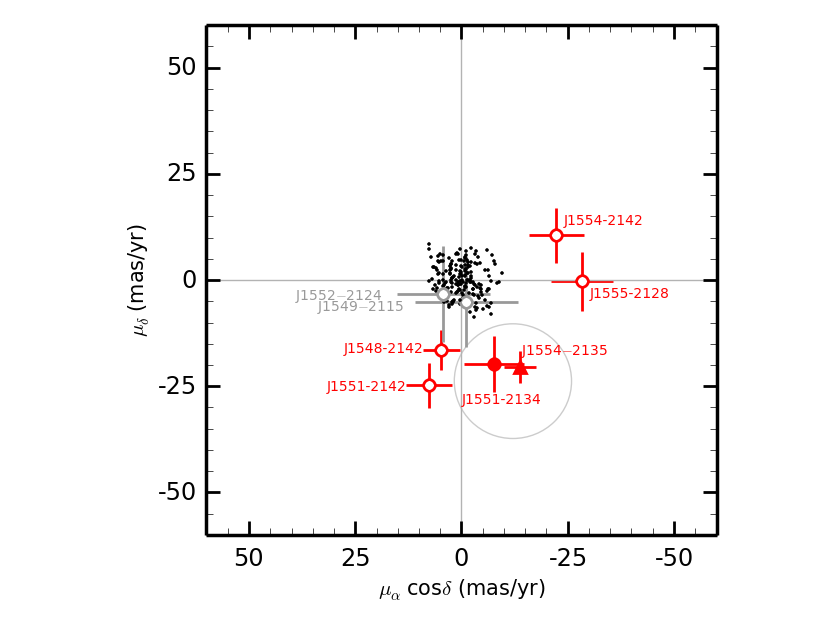}\\
\caption{Proper motion diagram of USco photometric member candidates. The seven photometric candidates are labelled and depicted in red color, while the two candidates rejected in Section~\ref{others} (also labelled) are shown in gray. The gray ellipse centered at ($-12.1$,$-23.8$) mas\,yr$^{-1}$ denotes our 2-$\sigma$ astrometric criterion to identify objects with proper motion consistent with USco membership, where $\sigma$ is the astrometric dispersion of USco low-mass members (see text). The sources used as astrometric references  
are presented with black dots. The high proper motion source 2MASS\,J15501151$-$2201213 (first discoverered by \citealt{deacon09}) lies off limits.
\label{upmfig}}
\end{figure}

As shown in Table~\ref{pmlos9}, for each USco candidate we managed to obtain a minimum of five proper motion measurements; all individual derivations are consistent with each other within the quoted uncertainties, except for a few cases where the S/N of the target is close to the detection limit. We adopted the weighted mean motions as the final values. 

Conservatively, the proper motion errors associated with the adopted motions for a given photometric candidate correspond to the largest uncertainty of the individual measurements where the target is detected with S/N greater than 10. Figure~\ref{upmfig} shows the proper motion diagram where the seven USco photometric candidates are depicted in red color. The expected location of USco members is given by the ellipse of semi-major axis of 13.8\,mas yr$^{-1}$ and semi-minor axis of 13.5\,mas yr centered at the motion of the young stellar association. The size of the ellipse was computed as twice the proper motion dispersions ($\pm$6.9 mas\,yr$^{-1}$ in $\mu_\alpha$, and $\pm$6.7 mas\,yr$^{-1}$ in $\mu_\delta$) observed among the USco low-mass members confirmed by \citet{lodieu08}; this is, the ellipse defines a 2-$\sigma$ criterion for the astrometric assessment of membership in USco. Five out 7 photometric candidates (J1554$-$2142, J1555$-$2128, J1548$-$2142, J1551$-$2142, and J1550$-$2201) lie outside the 2-$\sigma$ ellipse, indicating that they are not confimed members of USco by proper motion studies. The remaining two candidates have proper motions within the 2--$\sigma$ ellipse of Figure~\ref{upmfig}, thus confirming their likely membership in USco: J1551$-$2134 and J1554$-$2135. The latter was first identified by \citet{lodieu08}; their proper motion and our measurement agree within 0.2 $\sigma$. This number of confirmed members out of our original list of 7 potential USco candidates was expected from the level of field dwarf contamination discussed in Section~\ref{ucont}. 

\begin{figure}[h]
\centering
\includegraphics[width=0.35\textwidth]{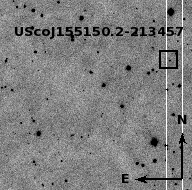}
\caption{Finding chart of the new USco member candidate J1551$-$2134. VIMOS $z$ band, 2 $\times$ 2\arcmin in size. 
\label{ucarta}}
\end{figure}

J1551$-$2134 is the new likely member of USco based on its photometry and proper motion found in the VIMOS--UKIDSS survey. Its motion deviates only by $\sim$0.3 $\sigma$ with respect to the mean proper motion of the stellar association. J1551$-$2134 shows optical and infrared colors that agree with a mid-L spectral classification. As indicated in Table~\ref{los12}, there is \textit{WISE} $W1$ and $W2$ photometry available for this object. Given the similarity between the \textit{Spitzer} $[4.5]$ and \textit{WISE} $W2$ magnitudes for L-type dwarfs \citep{zapatero11}, we investigated the presence of infrared flux excesses by means of color-color diagrams like those presented in Figure~5 of \citet{penaramirez12}. J1551$-$2134 does not appear to have significant color excesses up to 4.5 $\mu$m. On the contrary, this object has optical and infrared colors consistent with the indices of field, high-gravity L dwarfs \citep{kirkpatrick12}. Figure~\ref{ucarta} illustrates the finding chart of the new USco member.

For the sake of completeness, we also determined the proper motions of the two objects (J1552$-$2124 and J1549$-$2115) that were discarded as photometric candidates in Section~\ref{others}. The measurements are given in Table~\ref{pmlos9} and are plotted as gray circles in Figure~\ref{upmfig}. As expected, none has a proper motion compatible with the USco stellar association. 

%%%%%%%%%%%%%%%%%%%%%%%%%%%%%%%%%%%% 

\section{Spectroscopic follow--up of J1551$-$2134}\label{anal_spectrum}
\begin{figure}[]
\centering
\includegraphics[width=0.53\textwidth]{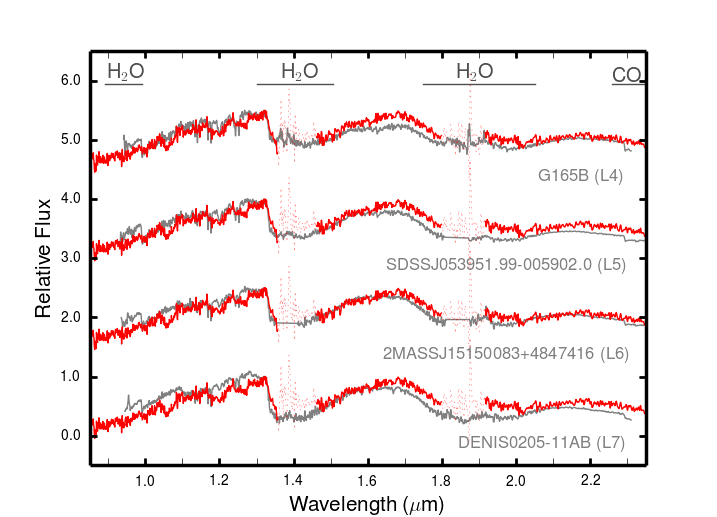}
\includegraphics[width=0.53\textwidth]{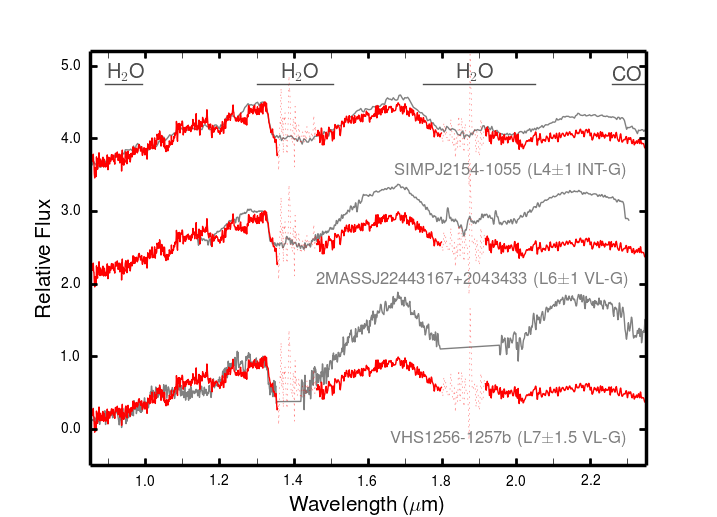}
\includegraphics[width=0.53\textwidth]{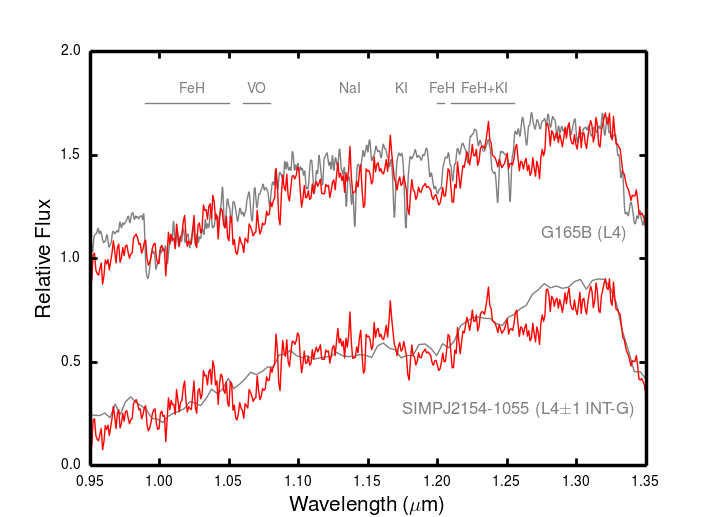}
\caption{FIRE low resolution near-infrared spectrum of J1551$-$2134 (solid red line) compared with field high-gravity dwarf templates (top panel) and with  young low- (VL--G) and intermediate-gravity (INT-G) L dwarfs (middle panel). The comparison spectra are labelled and plotted with solid gray lines (see text for proper references to the data). Red dotted lines depict the wavelength regions strongly affected by telluric absorption. The bottom panel illustrates the enlargement of J1551$-$2134's FIRE spectrum at around the K\,{\sc i} lines ($J$-band). All spectra are normalized to unity at 1.32 $\mu$m and are shifted by a constant in the vertical direction. Some molecular and atomic features are identified. 
\label{spectrum}}
\end{figure}

The FIRE spectrum of the new photometric and astrometric likely member, J1551$-$2134, is presented in Figure~\ref{spectrum}. It is compared with field, high gravity field L dwarfs in the top panel of the Figure, and with young, low gravity L dwarfs in the middle panel of the Figure. The comparison spectra were collected from different works: SDSS J053951.99$-$005902.0 and 2MASS J15150083$+$4847416 from \citet{rayner09}, SIMP J2154$-$1055 from  \cite{gagne14}, 2MASS J22443167$+$2043433, GD\,165B, and DENIS0205$-$11AB from \citet{mclean03}, and VHS J1256$-$1257b from \citet{gauza15}.  These data share a spectral resolution similar to the FIRE spectrum. As inferred from the top panel of Figure~\ref{spectrum}, J1551$-$2134 displays a red slope compatible with a spectral classification of L6 with an uncertainty of one subtype. This typing agrees with the optical and infrared colors of J1551$-$2134

At the age of the USco stellar association, substellar objects like J1551$-$2134 are expected to be undergoing gravitational self-contraction. Therefore, their atmospheres are governed by conditions of low pressure and low gravity. This impacts the fine details of the output flux at cool temperatures like those of the L types, particularly the atomic features: the lower the surface gravity, the weaker the atomic lines become (e.g., the Na\,{\sc i} doublet at 1.14\,$\mu$m, the two K\,{\sc i} doublets at 1.17 and 1.25\,$\mu$m). Also, the molecular features due to FeH (like the one at 1.19\,$\mu$m, which persists down to spectral type $\sim$L5) are affected. These and other signatures caused by low-pressure atmospheres are noticed in the $J$-band spectrum of J1551$-$2134. As illustrated in the bottom panel of Figure~\ref{spectrum}, the K\,{\sc i} lines are not detected in J1551$-$2134 at the resolution of our data. We determined an upper limit on the pseudo-equivalent widths of the $J$-band K\,{\sc i} lines of $\le$6 \AA. On the contrary, the comparison field, high-gravity dwarf (GD\,165B) has strong K\,{\sc i} absorptions. Furthermore, the 1.4--1.65\,$\mu$m H$_2$O features of the $H$-band adopt a ``triangular" shape due to low gravity atmospheres \citep{lucas01, kirkpatrick06}, a feature that is also seen in J1551$-$2134. Complementary, we also measured the gravity-sensitive indices defined for low-to-intermediate resolution spectra by \citet{allers13}, finding FeH$_z$ = 1.22\,$\pm$\,0.02,
FeH$_J$=1.16\,$\pm$\,0.03, K\,{\sc i}$_J$=0.94\,$\pm$\,0.01 and $H$-cont = 0.93\,$\pm$\,0.03  for J1551$-$2134. All indices are compatible with low-to-intermediate gravity scores. This result together with the upper limit on the strength of the $J$-band K\,{\sc i} lines and the ``triangular" shape of the $H$-band pseudo-continuum confirm the very low gravity nature of this object. The FIRE spectrum thus supports its membership in USco. J1551$-$2134 becomes a genuine young L6\,$\pm$\,1 object.

Although J1551$-$2134 displays clear spectral features of youth, the overall near-infrared spectral energy distribution is not similar to other low gravity L dwarfs of intermediate age (10--500\,Myr), which show very red $J-K$ colors, like those shown in the middle panel of Figure~\ref{spectrum}. On the contrary, the near-infrared colors and spectral slope of J1551$-$2134 resemble older (high gravity) field dwarfs and other similarly young mid-L dwarfs found in $\sigma$ Orionis (3 Myr) and USco \citep{martin01,lodieu08,penaramirez12,lodieu13surv,zapatero15}. This suggests that the very red $J-K$ colors observed from low gravity L dwarfs of intermediate age (typically $\ge$10 Myr) cannot be explained by the effects of low gravity atmospheres only, or at least the reddening of the $J-K$ color may depend nonlinearly on the low gravity. Alternative explanations such as scenarios based on warm dusty disks/envelopes have been proposed to reproduce the spectral behavior of very red L-type objects of intermediate age \citep{zapatero10,zakhozhay15}.

\section{Discussion}
\subsection{Masses of USco member candidates}\label{umodels}
To estimate the masses covered by our VIMOS--UKIDSS survey, we compared the $J$-band magnitudes of our photometric candidates with the evolutionary models by the Lyon group \citep{baraffe98,chabrier00model,baraffe03}. Theoretical luminosities and effective temperatures were converted into observables (magnitudes and colors) by applying bolometric correction--temperature and color--temperature relations given by \citet{golimowski04} and \citet{hewett06}. These relations are valid for field dwarfs, and we expect them to be also applicable to USco objects down to the mid-L types because of the similarity of the colors and spectral slopes between USco members and the field (see previous Sections). At the age of 5\,Myr and distance of 145\,pc, the substellar limit ($\sim$0.072\,M$_{\odot}$) is located at about $J$\,=\,13.0\,mag in the USco sequence of members, and the classical boundary between brown dwarfs and planets (0.012\,M$_{\odot}$ or $\sim$13\,M$_{\rm Jup}$) lies at $J$\,=\,16.6\,mag. The VIMOS--UKIDSS survey is complete in the mass interval that goes from 0.025\,M$_\odot$ through 0.004\,M$_\odot$ (corresponding to $J=14.5-19$\,mag). Down to the limiting magnitudes, the 1.17\,deg$^2$ survey would explore masses as small as 2--3\,\mj. These masses are labelled in Figures~\ref{ujzj}--\ref{udemas}. For the older age of 10\,Myr, our combined VIMOS--UKIDSS survey would be complete in the mass interval 0.005--0.028\,M$_\odot$, reaching down to 3--4\,M$_{\rm Jup}$ at the
limiting magnitudes. Either for 5\,Myr or 10\,Myr, our search was sensitive to the brown dwarf--planetary-mass object frontier and entered deeply in the planetary mass regime. 

Other published surveys \citep{lodieu06,lodieu11,bejar09,dawson13,dawson14} also covered the same 1.17 deg$^2$ area, which represents a modest 0.6$\%$ of the wide USco extension. These authors mainly employed data from the catalogs DENIS, 2MASS, and UKIDSS and uncovered vast extensions of this young stellar association. \citet{bejar09} and \citet{lodieu11} searched for the USco population within the mass interval $\approx$0.25--0.025\,\ms, while \citet{dawson13,dawson14} identified and studied spectroscopically USco members with masses ranging from 0.09\ms~to 0.01\ms~and spectral types M5--L1. Our VIMOS--UKIDSS survey, although smaller in area coverage, represents the extension towards lower planetary masses and later spectral types. The recent work by \citet{lodieu13surv}, with magnitude and mass sensitivities similar to our search, is the widest and deepest search for USco photometric candidates performed to date; however, it does not include the USco region studied by us.

As for the two photometrically, astrometrically, and spectroscopically confirmed USco members found in this work, we derived the following masses (5\,Myr, 145\,pc): $\approx$0.021\ms~(J1554$-$2135) and $\approx$8\mj (J1551$-$2134, new discovery). For an older age of 10\,Myr (see Section~\ref{intro}), the mass of the new USco member would turn to be $\approx$10\,$M_{\rm Jup}$. J1551$-$2134 appears to have a mass below the deuterium burning mass limit even if the USco association were 10\,Myr old; it becomes one of the least massive and late-type objects known in the entire USco association.

%%%%%%%%%%%%%%%%%%%%%%%%%%%%%%%%%%%% 

\subsection{Mass function} \label{final} 
We put our findings in the context of the mass function of young clusters and other star-forming regions. \citet{penaramirez12} presented the mass function of the 3\,Myr \so cluster that extends from the most massive stars (main sequence O-type star) all the way down to the planetary-mass domain at 4\,M$_{\rm Jup}$. These authors discussed that the \so cluster harbours about as many brown dwarfs (0.072--0.012\,M$_\odot$) and planetary-mass objects (0.012--0.004\,M$_\odot$) as low-mass stars (0.25--0.072\,M$_\odot$), with a mass spectrum that smoothly increases for low masses following the expression $\Delta N / \Delta M \sim M^{-\alpha}$, where $\alpha = 0.6\pm0.2$ for $M < 0.35\,$M$_\odot$. Other mass function derivations yielded rising functions with typical power-law indices between 0.4 and 1.0 in various star forming regions: ONC, $\rho$\,Oph, NGC2024, NGC1333, IC348, Cha\,I, Blanco 1, Pleiades, $\alpha$ Per, $\lambda$ Ori, NGC6611, USco, Lupus\,3. (See \citealt{bastian10,jeffries12} and \citealt{offner14} for summaries; and \citealt{muzic15} for a recent study).

Both USco and \so likely have related solar metallicity, and the two regions host massive stars, which allows us a direct comparison. In the VIMOS--UKIDSS area there are a total of 10 known USco member candidates with masses in the interval 0.18--0.025\,\ms~\citep{lodieu06,lodieu11,bejar09,dawson13,dawson14}, i.e., above the high-mass sensitivity limit of our survey. By considering the \so mass function of \citet{penaramirez12}, we predicted the presence of 3.5$^{+1.2}_{-1.0}$ USco members (using an age of 5\,Myr) with masses of 0.025--0.004\ms~equally distributed as follows: 1.7\,$\pm$\,0.4 objects for the mass bin 0.025--0.012\,\ms~and 1.8$^{+0.8}_{-0.6}$ sources populating the least massive interval 0.012--0.004\,\ms. A similar number and distribution of objects would be expected if the USco age were 10\,Myr. Despite being consistent with the expectations, the finding of two USco members in our survey favours the low values of the $\alpha$ exponent of the power law mass function better than the high values. What we indeed found was one USco member at each side of the brown dwarf---planetary-mass classical boundary.

As seen from the comparison of the USco string of members with the location of the field sequence of M, L, and T-type dwarfs shown in Figures~\ref{ujzj}--\ref{udemas}, our VIMOS--UKIDSS survey was designed deep enough to detect early-T and possibly mid-T type USco objects (this comparison did not account for the impact of low gravities on the spectral behaviour of the methane atmospheres). The evolutionary models suggest that USco potential members with temperatures below 1300 K (the L/T transition) and within the completeness magnitude of our survey would have masses in the interval 0.007--0.004\ms~($J$\,=\,18.57--19.00\,mag) The mass function of \citet{penaramirez12} predicts $\sim$0.8 objects of this kind. We found no candidates displaying colors typical of field T dwarfs, which is compatible with the predictions and allowed us to discard mass functions with $\alpha$\,$\ge$\,1.0, 1.1, and 1.2 with confidence levels of 90\%, 95\%, and 98\%, respectively.

\section{Summary and conclusions}\label{conc}
We used deep photometric $I$- and $z$-band data collected with the VIMOS instrument to perform a search for the least massive population of the young USco stellar association ($\sim$5--10\,Myr, 145\,pc). Combined with the UKIDSS catalog, the survey explored an area of 1.17 deg$^2$ (northeast of the extense USco region) in the magnitude and mass ranges $J=14.5-19$ mag and 0.028--0.004~\ms~(completeness). We also employed the {\sl WISE} catalog ($W1$ and $W2$ magnitudes) for the analysis of the photometric candidates. We found an initial list of 11 photometric $zJ$ candidates, which was later reduced to 7 after evaluation of the plethora of colors covering the wavelength interval 0.8--3.4 $\mu$m. The proper motion study confirmed only 2 USco members, one of which has a brown dwarf mass of $\approx$0.020--0.022~\ms~and was previously known, and the second object, J1551$-$2134, is a new discovery that has a planetary-mass of $\approx$0.008--0.010~\ms~and no apparent infrared flux excesses up to 4.5 $\mu$m. The near-infrared spectroscopic follow-up of J1551$-$2134 ($JHK$ FIRE spectrum of resolution 450 at 1.66 $\mu$m) confirmed the low-gravity nature of its atmosphere (weak alkaline lines, strong VO absorption, peaked $H$-band pseudocontinuum), as expected for a young cool source, and yielded a spectral type of L6\,$\pm$\,1. J1551$-$2134 shows optical and infrared colors resembling those of field, high gravity dwarfs and very young ($<$10\,Myr) members of similar classification in marked contrast with the very red indices of field, low-gravity L dwarfs of intermediate age. This suggests that gravity alone is not the key factor to account for the very red nature of some young L dwarfs and/or that the colors do not depend linearly on gravity. The finding of two USco substellar  
members in our VIMOS-UKIDSS survey is consistent with the low values of the exponent in the mass spectrum of $\sigma$ Orionis found by \citet{penaramirez12}. The non detection of T-type candidates in our survey allowed to constrain a mass spectrum in the interval 0.007--0.004\,M$_{\rm Jup}$, $\Delta N / \Delta M \sim M^{-\alpha}$, where $\alpha < 1.2$ with a confidence level of 98\%. J1551$-$2134 is one of the least massive and latest type members of the USco stellar association.

\begin{acknowledgements}
We thank the anonymous referee for providing us with constructive comments and suggestions. This work is based on observations made with the WHT, installed at the Spanish Observatorio del Roque de los Muchachos in the island of La Palma, Spain. Also based on observations made with ESO Telescopes at the Paranal Observatory under program ID 083.C-0556. This work is partly financed by the Spanish Ministry of Economy and Competitivity through the project AYA2014-54348-C3-2-R, and the Chilean FONDECYT Postdoctoral grant 3140351. \end{acknowledgements}

\bibliographystyle{aa} % style A&A
\bibliography{USco_paper}
%\nocite*
\end{document}